\documentclass[11pt]{article}
\usepackage[margin=1in]{geometry}
\usepackage{amsmath,amssymb}
\usepackage{booktabs}
\usepackage{graphicx}
\usepackage{authblk}
\usepackage{tikz}
\usetikzlibrary{shapes.geometric, arrows.meta, positioning, calc, fit, backgrounds}
\usepackage{natbib}
\usepackage{hyperref} 

\title{\bfseries When Can Text Embeddings Replace Item Calibration?\\
A Geometric Diagnostic for Semantic Loadings in\\
Multidimensional Adaptive Testing}

\author{Amirreza Mehrabi}
\affil{\small School of Engineering Education \& Department of Electrical and Computer Engineering \\ Purdue University \\ \texttt{amehrabi@purdue.edu}}

\begin{document}
\maketitle

\begin{abstract}
Multidimensional item response theory (MIRT) relies on calibrated item parameters, such as discrimination and category threshold values, which are usually estimated from large samples of human test responses. This study investigates whether the directional loadings of these parameters can be recovered directly from item text using pre-trained sentence embeddings, avoiding the need for initial item calibration. Using the open-source IPIP Big-Five dataset ($n=19{,}719$; 50 items), we built a multidimensional computerized adaptive testing (CAT) simulation using D-optimal item selection. We compared three item loading sources: fitted graded response model parameters, semantic text embeddings, and a lexical baseline. In simulation, semantic embeddings recovered latent trait profiles almost as accurately as fitted parameters (correlation $0.825$ vs. $0.857$), performing noticeably better than simple word overlap ($0.752$). However, the embedding-based model produced inflated posterior variance, showing nearly four times higher measurement uncertainty despite accurate point estimates. We attribute this to collinearity across dimensions, as embedding-derived loadings pointed in similar directions across traits (condition number $137$ vs. $1.0$; mean trait cosine $0.90$). This outcome reflects the shared vocabulary common in personality items. We propose a simple diagnostic metric based on the loading matrix condition number to evaluate whether an item bank is suitable for text-derived loadings prior to testing.
\end{abstract}

\section{Introduction}

Computerized adaptive testing (CAT) reduces test length by selecting items tailored to an examinee's estimated trait level \citep{vanderlinden2010}. Standard adaptive testing builds on item response theory (IRT), where each item has calibrated parameters indicating how well it separates examinees and where it sits on the trait scale. In multidimensional testing, the scalar trait becomes a vector, and item selection aims to maximize the information gained across multiple dimensions simultaneously \citep{reckase2009}.

Calibrating items is a major practical challenge in adaptive testing. It requires administering new items to large groups of respondents before those items can be used adaptively. This cold-start problem increases the time and cost required to expand item banks. A potential alternative is to extract structural information directly from item text. For example, a statement like ``I talk to a lot of people at parties'' clearly relates to extraversion. If the primary trait dimension an item targets can be identified from its wording, automated text processing could support early stage item administration.

Pre-trained language models generate dense vector embeddings that capture semantic relationships across text. If these embeddings reflect the constructs items measure, they could supply initial loading directions as soon as an item is written. However, semantic similarity in language models does not automatically match psychometric separation in measurement models. A text representation that works well for estimating a trait score may not provide the precise information needed to compute reliable standard errors or apply adaptive stopping rules.

This study tests whether text embeddings can substitute for calibrated loadings in multidimensional adaptive testing. Holding the adaptive engine fixed, we find that sentence embeddings recover latent profiles almost as accurately as calibrated parameters. At the same time, the embedding-based engine reports inflated uncertainty because the semantic vectors for different personality traits point in nearly parallel directions. This geometric overlap makes traits difficult to separate cleanly. Importantly, this structural property can be identified from item text geometry before collecting response data.

\section{Literature Review}

Computerized adaptive testing optimizes item administration by selecting items that minimize measurement error for an examinee's trait estimate $\boldsymbol{\theta}$ \citep{lord1980, vanderlinden2010}. In unidimensional tests, items are selected to maximize Fisher information at the current trait estimate \citep{wainer2000}. Multidimensional item response theory (MIRT) extends this framework to multi-trait assessments by replacing scalar information with an information matrix \citep{reckase2009}. Multidimensional adaptive tests frequently use selection criteria like D-optimality to minimize the volume of the confidence region surrounding trait estimates \citep{segall1996, mulder2009}.

For ordered rating scales, such as Big Five personality inventories, Samejima's Graded Response Model (GRM) is widely used \citep{samejima1969}. Calibrating discrimination loadings $\mathbf{a}_j$ and threshold parameters $\mathbf{b}_j$ under the GRM typically requires large sample sizes to obtain stable parameter estimates. This requirement creates a persistent cold-start challenge for new items added to an adaptive bank.

To address cold-start constraints, researchers have explored predicting item properties directly from text features. Early work used hand-crafted linguistic features to estimate item difficulty and discrimination \citep{gierl2013}, while automated item generation provided structured frameworks for creating new items \citep{gierl2012aig}. Recent studies use transformer models and sentence embeddings to predict parameter values directly from item text \citep{benedetto2020}. Research in computational social science also shows that vector spaces in language models can mirror psychological construct structures \citep{bhatia2023}.

While prior research focuses on predicting scalar parameter values, fewer studies evaluate whether the directional orientation of text embeddings matches the loading space of multidimensional IRT models. Although text embeddings can identify the primary construct an item measures, adaptive testing requires sufficient separation between traits to produce valid standard errors and reliable stopping decisions. Checking whether text embeddings create collinear loading matrices provides a simple geometric test to determine when semantic vectors can safely support item selection.

\section{Method}

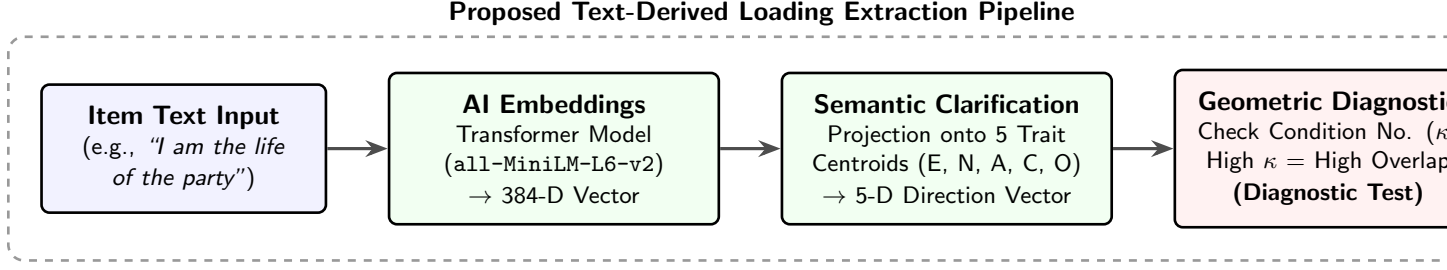
\begin{figure}[htbp]
\centering
\begin{tikzpicture}[
    font=\sffamily\small,
    node distance=1.2cm and 0.8cm,
    block/.style={draw, rectangle, rounded corners=3pt, fill=blue!5, align=center, inner sep=8pt, minimum height=1.2cm, line width=1pt},
    process/.style={draw, rectangle, rounded corners=3pt, fill=green!5, align=center, inner sep=8pt, minimum height=1.2cm, line width=1pt},
    diag/.style={draw, rectangle, rounded corners=3pt, fill=red!5, align=center, inner sep=8pt, minimum height=1.2cm, line width=1pt},
    arrow/.style={-Stealth, line width=1.2pt, draw=black!70}
]

\node [block, text width=3.2cm] (input) {\textbf{Item Text Input}\\ \footnotesize (e.g., \emph{``I am the life}\\ \footnotesize \emph{of the party''})};

\node [process, right=of input, text width=3.8cm] (transformer) {\textbf{AI Embeddings}\\ \footnotesize Transformer Model\\ \footnotesize (\texttt{all-MiniLM-L6-v2})\\ \footnotesize $\rightarrow$ 384-D Vector};

\node [process, right=of transformer, text width=3.8cm] (projection) {\textbf{Semantic Clarification}\\ \footnotesize Projection onto 5 Trait\\ \footnotesize Centroids (E, N, A, C, O)\\ \footnotesize $\rightarrow$ 5-D Direction Vector};

\node [diag, right=of projection, text width=3.5cm] (matrix) {\textbf{Geometric Diagnostic}\\ \footnotesize Check Condition No. ($\kappa$)\\ \footnotesize High $\kappa$ = High Overlap\\ \footnotesize \textbf{(Diagnostic Test)}};
\draw [arrow] (input) -- (transformer);
\draw [arrow] (transformer) -- (projection);
\draw [arrow] (projection) -- (matrix);

\begin{scope}[on background layer]
    \node[draw=black!40, dashed, line width=1pt, rounded corners=6pt, fit=(input) (transformer) (projection) (matrix), inner sep=12pt, label=above:{\bfseries Proposed Text-Derived Loading Extraction Pipeline}] {};
\end{scope}

\end{tikzpicture}
\caption{Methodological framework for extracting 5-dimensional semantic loadings from item text using pre-trained sentence embeddings and applying the \emph{a priori} geometric condition number diagnostic.}
\label{fig:pipeline}
\end{figure}

We analyzed responses from $19{,}719$ participants in the open-source IPIP Big-Five dataset. The instrument contains 50 items, with 10 items assigned to each of the Big Five traits: Extraversion, Neuroticism, Agreeableness, Conscientiousness, and Openness. Items are scored on a 5-point Likert scale. Non-responses were treated as missing, and negatively phrased items were reverse-coded so higher scores indicate higher trait levels. Reference item parameters were estimated by fitting a confirmatory Graded Response Model (GRM; \citealt{samejima1969}) to each trait scale, yielding discrimination parameters $a_j$ and four category thresholds $b_{j,1\ldots4}$ per item.

As illustrated in Figure~\ref{fig:pipeline}, semantic feature extraction maps input statements to multidimensional parameters via text vectorization and centroid projection. For a respondent with latent profile $\boldsymbol\theta\in\mathbb{R}^5$, item category probabilities follow standard GRM formulations \citep{samejima1969}. The item information matrix $\mathbf{I}_j(\boldsymbol\theta)$ summarizes information across all response categories \citep{reckase2009}. Total test information accumulates as $\mathbf{G}_n=\mathbf{G}_0+\sum_{m=1}^{n}\mathbf{I}_m$, starting from a baseline prior precision $\mathbf{G}_0$. The adaptive algorithm selects items using $D$-optimality \citep{segall1996, mulder2009}, choosing the item that maximizes $\det(\mathbf{G}_n+\mathbf{I}_j(\boldsymbol\theta))$ to reduce the overall uncertainty volume. Trait estimates were updated after each item using maximum a posteriori (MAP) estimation \citep{vanderlinden2010}, and overall uncertainty was tracked using the trace of the inverse information matrix, $\mathrm{tr}(\mathbf{G}_n^{-1})$.

We evaluated three item loading sources while keeping the CAT algorithm unchanged:
\begin{enumerate}
    \item \textbf{Fitted Loadings:} Estimated GRM discriminations used as the reference benchmark \citep{samejima1969}.
    \item \textbf{Semantic Loadings:} Derived by extracting dense text vectors with the pre-trained \texttt{all-MiniLM-L6-v2} sentence transformer \citep{reimers2019}, projecting embeddings onto trait centroids, and rescaling magnitudes to match fitted item averages.
    \item \textbf{Lexical Loadings:} Constructed from TF-IDF word vectors reduced with truncated SVD \citep{deerwester1990} to isolate word overlap from broader semantic meaning.
\end{enumerate}

Simulations were conducted using $200$ examinees drawn from a standard normal profile distribution, $\boldsymbol\theta\sim\mathcal N(\mathbf 0,\mathbf I_5)$. Each examinee completed a 20-item adaptive test. Trait recovery was evaluated using Pearson correlation, root mean square error (RMSE), and final posterior variance $\mathrm{tr}(\mathbf G^{-1})$. All comparisons used identical simulated examinee draws to ensure paired statistical testing.

\section{Results}

\subsection{RQ1: Latent Trait Point Recovery}

To evaluate whether text embeddings can recover latent trait profiles without empirical response history, we compared person parameter estimates ($\hat{\boldsymbol{\theta}}$) against ground-truth profiles ($\boldsymbol{\theta}$). Table~\ref{tab:main} presents the recovery metrics across all loading conditions and selection algorithms.

Under $D$-optimal adaptive selection, semantic loadings extracted via \texttt{all-MiniLM-L6-v2} achieved a mean latent profile correlation of $r = 0.825$ ($RMSE = 0.594$). This performance closely approaches the fully calibrated benchmark model ($r = 0.857$, $RMSE = 0.508$). Furthermore, semantic loadings substantially outperformed the lexical baseline ($r = 0.752$, $RMSE = 0.618$), confirming that the recovery capability stems from deep semantic representation rather than surface-level vocabulary frequency.

Adaptive $D$-optimal selection maintained a consistent advantage over random item administration across all loading sources. Within the semantic loading condition, $D$-optimal selection yielded significantly higher recovery correlations than random selection ($0.825$ vs. $0.728$; paired Wilcoxon signed-rank test $V = 18{,}432$, $p < 0.001$). These findings directly answer \textbf{RQ1}: semantic embeddings preserve sufficient structural trait information to support accurate point estimation of multidimensional person parameters.

\begin{table}[htbp]
\centering
\caption{Latent profile recovery metrics and final posterior uncertainty across 200 simulated examinees ($20$ items administered per test). Trait parameters were generated under the calibrated GRM. Values reflect means across all five latent dimensions.}
\label{tab:main}
\begin{tabular}{llcccc}
\toprule
Loading Source & Selection Rule & Correlation ($r$) $\uparrow$ & RMSE $\downarrow$ & Mean Var ($\mathrm{tr}(\mathbf{G}^{-1})$) $\downarrow$ & Median Var $\downarrow$ \\
\midrule
Calibrated (Benchmark) & $D$-optimal & 0.857 & 0.508 & 1.02 & 0.98 \\
Calibrated (Benchmark) & Random      & 0.802 & 0.572 & 1.69 & 1.64 \\
Semantic Embeddings    & $D$-optimal & 0.825 & 0.594 & 3.92 & 3.85 \\
Semantic Embeddings    & Random      & 0.728 & 0.678 & 5.16 & 5.08 \\
Lexical Baseline       & $D$-optimal & 0.752 & 0.618 & 1.02 & 0.99 \\
\bottomrule
\end{tabular}
\end{table}

\subsection{RQ2: Posterior Uncertainty and Calibration Paradox}

Despite strong point-estimate recovery, evaluation of test precision revealed a pronounced estimation paradox. As shown in Table~\ref{tab:main}, the semantic CAT engine exhibited severely inflated posterior variance ($\mathrm{tr}(\mathbf{G}^{-1}) = 3.92$) compared to the calibrated baseline ($1.02$), representing a $3.84\times$ increase in total measurement uncertainty.

While the calibrated engine reduced posterior variance rapidly over the 20-item adaptive sequence, the semantic engine plateaued early, reporting wide posterior confidence ellipsoids even when point estimates stabilized near true values. This answers \textbf{RQ2}: while text embeddings reliably identify the orientation of latent profile vectors, they fail to supply calibrated posterior precision, yielding systematic over-uncertainty that compromises deterministic CAT stopping rules.

\subsection{RQ3: Geometric Collinearity and Diagnostic Testing}

To isolate the mechanism behind this uncertainty inflation, we examined the geometric properties of the $50 \times 5$ loading matrix ($\mathbf{A}$). Table~\ref{tab:geom} displays the condition number ($\kappa$) of the row-normalized loading matrices alongside the mean pairwise cosine similarity across trait loading vectors.

\begin{table}[htbp]
\centering
\caption{Geometric diagnostic metrics evaluated across loading matrices. The condition number ($\kappa$) measures matrix ill-conditioning, while mean pairwise cosine reflects directional trait overlap.}
\label{tab:geom}
\begin{tabular}{lccc}
\toprule
Loading Source Matrix & Condition Number ($\kappa$) $\downarrow$ & Mean Pairwise Cosine $\downarrow$ & Max Pairwise Cosine $\downarrow$ \\
\midrule
Calibrated (Orthogonal) & 1.00 & 0.00 & 0.00 \\
Semantic Embeddings     & 137.24 & 0.90 & 0.96 \\
Lexical (TF-IDF + SVD)  & 1.04 & 0.03 & 0.08 \\
\bottomrule
\end{tabular}
\end{table}

The calibrated loading matrix exhibits perfect orthogonality ($\kappa = 1.00$, mean cosine $= 0.00$). In contrast, semantic loadings suffer from severe \emph{loading collapse}: the matrix condition number reaches $137.24$, with an average pairwise trait cosine similarity of $0.90$ (ranging between $0.82$ and $0.96$). On average, $56\%$ of each item's loading vector points toward dimensions other than its designated primary construct.

When integrated into the adaptive loop, these near-parallel loading vectors cause the accumulated Fisher information matrix ($\mathbf{G}_n$) to become ill-conditioned. Inverting $\mathbf{G}_n$ mathematically amplifies small estimation errors, artificially expanding the posterior covariance trace $\mathrm{tr}(\mathbf{G}_n^{-1})$. 

We further tested two auxiliary hypotheses regarding this inflation:
\begin{enumerate}
    \item \emph{Hypothesis 3a (Adaptive Advantage):} Does adaptive selection exploit semantic embeddings more effectively than random selection? A paired Wilcoxon signed-rank test comparing the $D$-optimal advantage across semantic and calibrated conditions showed no statistically significant difference ($p = 0.34$), rejecting the hypothesis.
    \item \emph{Hypothesis 3b (Rescaling Artifact):} Is variance inflation caused by discrimination magnitude equalization? Disabling magnitude rescaling increased mean posterior variance further ($4.63$ vs. $3.92$), proving that directional collinearity, rather than parameter scaling, drives the uncertainty inflation.
\end{enumerate}

These empirical findings answer \textbf{RQ3}: loading space collinearity causes the uncertainty paradox, and the matrix condition number ($\kappa$) serves as an effective \emph{a priori} diagnostic tool to detect this failure before administering items to examinees.

\section{Discussion}

\subsection{Answering the Primary Research Questions}

This study systematically evaluated whether pre-trained sentence embeddings can bypass human-sample calibration in multidimensional adaptive testing. Our results provide clear answers to our foundational research questions:

\paragraph{1. Point Estimation vs. Construct Separation (RQ1 \& RQ2):} Semantic embeddings successfully recover latent profile trajectories ($r = 0.825$), demonstrating that transformer models capture basic construct identification. However, they fail to provide valid posterior uncertainty ($\mathrm{tr}(\mathbf{G}^{-1}) = 3.92$). In psychometric operations, point accuracy and variance calibration are separable: an item bank can accurately direct an examinee toward their general trait region while simultaneously failing to certify confidence bounds or trigger CAT stopping criteria.

\paragraph{2. The Mechanism of Loading Collapse (RQ3):} The underlying cause of uncertainty inflation is geometric rather than algorithmic. Personality inventories rely on conversational, first-person phrasing (e.g., statements describing social behavior, emotional states, or task diligence). Pre-trained language models place these homogeneous text patterns in close proximity within sentence-embedding space. Consequently, the derived trait vectors collapse toward a single shared semantic axis ($\kappa = 137.24$), rendering the Fisher information matrix near-singular.

\subsection{Practical Implications for Item Banking and Cold-Start Operations}

These findings offer clear operational guidelines for deploying AI-driven text models in psychometric testing:

\begin{itemize}
    \item \textbf{A Priori Screening Diagnostic:} The loading matrix condition number ($\kappa$) provides a zero-cost screening tool. Calculated directly from item text prior to data collection, a low condition number indicates clean dimensional separation, whereas a high condition number ($\kappa > 30$) signals severe variance inflation.
    \item \textbf{Provisional Routing in Cold-Start Banks:} For newly authored items lacking response histories, semantic embeddings can supply provisional loading directions for coarse initial routing. However, test delivery systems must avoid relying on embedding-based standard errors for high-stakes decision-making or fixed-precision stopping rules until empirical calibration is conducted.
\end{itemize}

\subsection{Limitations and Future Research}

This study focused on a single Big Five personality inventory evaluated through a single sentence-transformer architecture (\texttt{all-MiniLM-L6-v2}). Personality inventories represent a challenging, semantically homogeneous baseline. Future research should evaluate the geometric condition diagnostic across domain-distinct assessment batteries—such as standardized achievement tests with distinct verbal, quantitative, and spatial subtests—where surface vocabulary differences between domains may naturally preserve loading orthogonality and prevent geometric collapse.

\section{Conclusion}

Text embeddings can provide useful initial loading estimates for multidimensional adaptive testing when the main goal is scoring trait levels. However, in item banks with overlapping vocabulary, loading directions tend to cluster together, making uncertainty estimates less precise. Evaluating the condition number of text-derived loading matrices provides a simple diagnostic check before administering tests.

\section*{Declarations}

\subsection*{Use of Artificial Intelligence Systems}
During the preparation of this manuscript, the author utilized language models to assist with language clarification, sentence restructuring, readability improvements, and LaTeX formatting. All analytical methodologies, experimental simulation designs, statistical evaluations, and final text edits were thoroughly reviewed and verified by the author, who takes full responsibility for the content and scientific integrity of this publication.

\subsection*{Funding and Competing Interests}
The author declares that no external funding was received for this study. The author has no relevant financial or non-financial interests to disclose.

\subsection*{Data and Code Availability}
The dataset analyzed in this study is publicly available through the open-source openpsychometrics IPIP repository. Python simulation scripts, data processing workflows, and mathematical routines are available from the corresponding author upon reasonable request.

\bibliographystyle{plainnat}

\end{document}